# Disc novae: thermodynamics of gas-assisted binary black hole formation in AGN discs

Henry Whitehead [ORCID],[1]★ Connar Rowan [ORCID],[2] Tjarda Boekholt [ORCID][2,3] and Bence Kocsis [ORCID][2,4]

[1]*Astrophysics Sub-department, Department of Physics, University of Oxford, Denys Wilkinson Building, Keble Road, Oxford OX1 3RH , UK*
[2]*Rudolf Peierls Centre for Theoretical Physics, Clarendon Laboratory, University of Oxford, Parks Road, Oxford OX1 3PU, UK*
[3]*NASA Ames Research Center, Moffett Field, CA 94035, USA*
[4]*St Hugh's College, St Margaret's Road, Oxford OX2 6LE, UK*

## ABSTRACT

We investigate the thermodynamics of close encounters between stellar mass black holes (BHs) in the gaseous discs of active galactic nuclei (AGNs), during which binary black holes (BBHs) may form. We consider a suite of 2D viscous hydrodynamical simulations within a shearing box prescription using the Eulerian grid code ATHENA++. We study formation scenarios where the fluid is either an isothermal gas or an adiabatic mixture of gas and radiation in local thermal equilibrium. We include the effects of viscous and shock heating, as well as optically thick cooling. We co-evolve the embedded BHs with the gas, keeping track of the energetic dissipation and torquing of the BBH by gas and inertial forces. We find that compared to the isothermal case, the minidiscs formed around each BH are significantly hotter and more diffuse, though BBH formation is still efficient. We observe massive blast waves arising from collisions between the radiative minidiscs during both the initial close encounter and subsequent periapsis periods for successfully bound BBHs. These 'disc novae' have a profound effect, depleting the BBH Hill sphere of gas and injecting energy into the surrounding medium. In analysing the thermal emission from these events, we observe periodic peaks in local luminosity associated with close encounters/periapses, with emission peaking in the optical/near-infrared (IR). In the AGN outskirts, these outbursts can reach 4 per cent of the AGN luminosity in the IR band, with flares rising over 0.5–1 yr. Collisions in different disc regions, or when treated in 3D with magnetism, may produce more prominent flares.

## 1 INTRODUCTION

Binary black holes (BBHs) are the dominant source of gravitational waves (GWs) as detected by terrestrial observatories LIGO (Laser Interferometer Gravitational-Wave Observatory)/Virgo/KAGRA (Kamioka Gravitational Wave Detector) (Abbott et al. 2016, 2019, 2020a,b,c,d, 2022).The origin of these binary mergers remains an open question (see Tagawa, Haiman & Kocsis 2020a and references therein). Recently, BBHs embedded within the discs of active galactic nuclei (AGNs) have received significant attention (McKernan et al. 2012, 2014, 2020; Stone, Metzger & Haiman 2017; Secunda et al. 2019; Yang et al. 2019; Gröbner et al. 2020; Tagawa et al. 2020a,b, 2021a,b). AGNs provide fertile ground for BBH formation due to the expected high density of compact objects in the galactic core (Bahcall & Wolf 1976; Miralda-Escudé & Gould 2000) and the ability for gas dynamical friction to align black holes (BHs) with the disc (Bartos et al. 2017; Panamarev et al. 2018) and for gas torques within the disc to form migration traps, leading to overdensities of approximately co-planar BHs (McKernan et al. 2012; Bellovary et al. 2016). Most importantly, the presence of gas in the AGN disc provides a dissipative medium to capture initially unbound BHs into binaries, either by dynamical drag through gas gravitation (Ostriker 1999) or by accretion of gas momentum. This results in a capture cross-section significantly larger than that by GW emission alone (O'Leary, Kocsis & Loeb 2009).

AGN embedded binaries have promise as a potential source of GW-EM (electromagnetic) multimessenger detections due to the gas-rich environment surrounding the binary. The mechanism producing EM emission coincident with the merger of these binaries varies between models (McKernan et al. 2019; Graham et al. 2020; Kimura, Murase & Bartos 2021; Wang et al. 2021; Tagawa et al. 2023b). Recently, Tagawa et al. (2023a) considered the EM signature from a jet generated by a single embedded BH. We are unaware of previous work considering an EM counterpart to binary formation itself, the focus of this work.

Much work has been done in understanding the behaviour and long-term evolution of BBHs with circumbinary discs (Tang, MacFadyen & Haiman 2017; Moody, Shi & Stone 2019; Muñoz, Miranda & Lai 2019; Duffell et al. 2020; Heath & Nixon 2020; Tiede et al. 2020; Westernacher-Schneider et al. 2022; Valli et al. 2024), as well as the more specific pre-existing AGN-embedded BBH system (Baruteau, Cuadra & Lin 2011; Li & Lai 2022, 2023, 2024). These studies are key to improving our understanding of how an initially soft binary can harden sufficiently to reach the separations where GW emission can drive inspiral to merger.

This work follows many recent studies into the formation of BBHs embedded in gaseous systems. These studies have spanned a variety of different numerical methods, including semi-analytical gas dynamical friction (DeLaurentiis, Epstein-Martin & Haiman 2023; Qian, Li & Lai 2024; Rozner, Generozov & Perets 2023), smoothed-particle hydrodynamics (Rowan et al. 2023, 2024, henceforth CR1 and CR2, respectively), and adaptive mesh Eulerian grid codes (Li et al. 2023; Whitehead et al. 2024). Our previous study (Whitehead et al. 2024, henceforth HW1) analysed a series of different BH trajectories and ambient disc densities in an isothermal shearing box prescription.

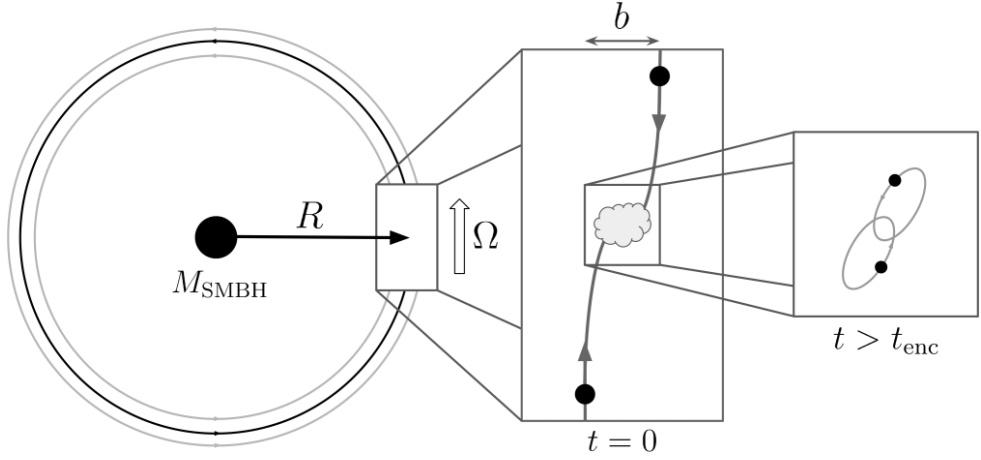

**Figure 1.** Schematic layout for the shearing box prescription. Only a Cartesian patch of the global AGN disc is simulated, in a frame that co-rotates with the local disc at the Keplerian rate $\Omega^2 = GM_{\rm SMBH}/R^3$. The BHs are initialized at $t = 0$ on circular Keplerian orbits with a radial separation of $b$; they will undergo an encounter within the shearing frame and will form an eccentric binary if sufficient orbital energy is dissipated during their encounter at $t = t_{\rm enc}$.

We expect the choice of equation of state to be important for BBH capture, as it may affect the formation and morphology of BH minidiscs and change the flow structure during and after the BH close encounters. Previous studies of embedded binary formation have used an isothermal equation of state (in part to reduce computational expense), but this choice may not faithfully represent a realistic capture environment. Studies considering the effect of utilizing a non-isothermal barotropic fluid with a pre-existing binary system (Li & Lai 2022, 2023, 2024) found significant morphological differences when the adiabatic exponent $\gamma$ was varied: the circumsingle discs tended to be hotter and less massive for systems with larger $\gamma$. This paper improves on previous binary formation models in isothermal discs and pre-existing binary evolution studies in barotropic gas environments by considering binary formation with a more realistic equation of state.

In this paper, we study the interactions between two initially unbound BHs embedded within an AGN disc. These interactions are studied using 2D viscous hydrodynamical simulations, with either an isothermal equation of state or an adiabatic mixture of gas and radiation in local thermal equilibrium. We analyse a suite of 38 simulations, examining how different BH trajectories and changes to the thermodynamics influence the fluid evolution and subsequently affect the likelihood of successful BBH formation. Section 2 presents our computational methodology, and the initial conditions are described in Section 3. In Section 4, we present the results and discuss novel phenomena that arise in simulations with a radiative equation of state and their possible detectability. In Sections 5 and 6, we discuss our findings and summarize our conclusions.

## 2 COMPUTATIONAL METHODS

This work builds directly on HW1, with much of the computational set-up maintained; we give a brief overview here for clarity. We use the Eulerian GRMHD (General Relativistic Magnetohydrodynamics) code ATHENA++ (Stone et al. 2020) to perform our hydrodynamical simulations, but neglect any effects associated with gas self-gravity, relativity, and magnetism. We utilize a second-order accurate van Leer predictor–corrector integrator with a piecewise linear method spatial reconstruction and Roe's linearized Riemann solver. Our simulation tracks a 2D rectangular patch of disc (shearing box) co-rotating around the SMBH (Supermassive Black Hole) with Keplerian angular frequency $\Omega$ (see Fig. 1). Within this box, our BHs function as point masses that are self-consistently propagated through the flow. The natural length-scales in the shearing box are the Hill radii of a single BH or BBH ($r_{\rm H,s}$ and $r_{\rm H}$, respectively):

$$r_{\rm H,s} = R\left(\frac{m_{\rm BH}}{3M_{\rm SMBH}}\right)^{1/3}, \quad (1)$$

$$r_{\rm H} = R\left(\frac{M_{\rm bin}}{3M_{\rm SMBH}}\right)^{1/3}, \quad (2)$$

where $m_{\rm BH}$, $M_{\rm bin}$, and $M_{\rm SMBH}$ are the mass of a single BH, the BBH, and the central SMBH, respectively, and $R$ is the separation between the shearing frame centre and the SMBH.

### 2.1 Gas dynamics

Gas within the shearing box evolves according to the extended Navier–Stokes equations, which include contributions from BH gravitation and inertial forces from the box's acceleration.

$$\frac{\partial \rho}{\partial t} + \nabla \cdot (\rho \mathbf{u}) = 0, \quad (3)$$

$$\frac{\partial (\rho \mathbf{u})}{\partial t} + \nabla \cdot (\rho \mathbf{u}\mathbf{u} + P\mathbf{I} + \mathbf{\Pi}) = \rho\left(\mathbf{a}_{\rm SMBH} + \mathbf{a}_{\rm BH}\right), \quad (4)$$

where we have introduced $\rho$, $P$, $\mathbf{u}$, and $\mathbf{\Pi}$ as the gas density, pressure, velocity, and viscous stress tensor

$$\Pi_{ij} = \rho \nu \left(\frac{\partial u_i}{\partial x_j} + \frac{\partial u_j}{\partial x_i} - \frac{2}{3}\delta_{ij}\nabla \cdot \mathbf{u}\right) \quad (5)$$

for a given kinematic viscosity $\nu$ (see Section 2.1.4).

#### 2.1.1 Shearing forces and gravity

We add to the Navier–Stokes equations a term $\mathbf{a}_{\rm SMBH}$ to account for acceleration of gas due to the central SMBH inducing a rotating frame. This acceleration can be represented in terms of the angular velocity of the frame $\Omega$ and background shear parameter $q = \frac{d\ln\Omega}{d\ln R} = \frac{3}{2}$ for Keplerian rotation:

$$a_{\rm SMBH} = 2\boldsymbol{u} \times \Omega\hat{\boldsymbol{z}} + 2q\Omega^2 \boldsymbol{x} \,. \tag{6}$$

In this frame, there exist equilibrium trajectories $\boldsymbol{u}_{\rm eq}$ for which $\boldsymbol{a}_{\rm SMBH} = 0$, with radial and azimuthal components

$$\boldsymbol{u}_{\rm eq} = \begin{pmatrix} 0 \\ -q\Omega x \end{pmatrix}. \tag{7}$$

These trajectories describe particles on circular Keplerian orbits around the SMBH at different orbital radii relative to the centre of the frame, and so form the basis trajectories for BHs launched into the simulation. BH gravitation is included in the simulation as

$$\boldsymbol{a}_{\rm BH} = -\nabla \phi_{\rm BH}(\boldsymbol{r}) = \sum_{n=1}^{n_{\rm BH}} m_{{\rm BH},n}\, g\left(\frac{\boldsymbol{r} - \boldsymbol{r}_n}{h}\right), \tag{8}$$

where $h$ is the softening length for the gravitational spline kernel $g(\delta)$ (Price & Monaghan 2007, Appendix A).

$$g(\delta) = -\frac{G}{h^2}\hat{\delta} \begin{cases} \frac{32}{3}\delta - \frac{192}{5}\delta^3 + 32\delta^4 & 0 < \delta \leq \frac{1}{2} \\ -\frac{1}{15\delta^2} + \frac{64}{3}\delta - 48\delta^2 + \frac{192}{5}\delta^3 - \frac{32}{3}\delta^4 & \frac{1}{2} < \delta \leq 1 \\ \frac{1}{\delta^2} & \delta > 1 \end{cases} \tag{9}$$

Softening is only applied to the BH–gas interactions: the BH–BH gravitation is unsoftened.

### 2.1.2 Equation of state

For non-isothermal equations of state, we must solve the energy equation

$$\frac{\partial E}{\partial t} + \nabla \cdot [(E + P)\boldsymbol{u} + \boldsymbol{\Pi} \cdot \boldsymbol{u}] = \rho \boldsymbol{u} \cdot (\boldsymbol{a}_{\rm SMBH} - \boldsymbol{a}_{\rm BH}) - Q. \tag{10}$$

Here, $E$ represents the total fluid energy per unit volume, separable into internal and kinetic components.

$$E = U + K = U + \frac{1}{2}\rho \boldsymbol{u} \cdot \boldsymbol{u}. \tag{11}$$

The internal energy density $U$ is dependent on the choice of equation of state (EoS). In this study, we consider an adiabatic, radiative EoS where gas and radiation contribute to the local pressure and energy density. We compute the total pressure and internal energy density as

$$P = P_g + P_r = \frac{k_{\rm B}}{\mu_p m_u}\rho T + \frac{1}{3}a_r T^4, \tag{12}$$

$$U = U_g + U_r = \frac{3}{2}\frac{k_{\rm B}}{\mu_p m_u}\rho T + a_r T^4, \tag{13}$$

where $k_{\rm B}$, $\mu_p$, $m_u$, and $a_r$ are the Boltzmann constant, average molecular weight, atomic mass constant, and radiation constant, respectively. Assuming for simplicity that gas is dominated by fully ionized hydrogen, we set $\mu_p = 0.5$. For a gas-radiation mixture, the adiabatic sound speed $a_s$ can be calculated via $\Gamma_1$, the first adiabatic exponent (Chandrasekhar 1939) as follows.

$$a_s^2 = \Gamma_1 \frac{P}{\rho}, \quad \text{where } \Gamma_1 = \frac{32 - 24\beta - 3\beta^2}{24 - 21\beta}. \tag{14}$$

Here, $\beta$ is the gas pressure fraction $\beta = \frac{P_g}{P}$. In the gas dominated ($\beta \to 1$) and radiation dominated ($\beta \to 0$) limits, $\Gamma_1$ evaluates to $\frac{5}{3}$ and $\frac{4}{3}$, respectively.

ATHENA++ supports the use of a general equation of state provided mappings between conserved and primitive variables are defined (Coleman 2020):

$$P = f_1(\rho, E) \tag{15}$$

$$E = f_2(\rho, P) \tag{16}$$

$$a_s = f_3(\rho, P) \tag{17}$$

Unlike for the trivial case of an isothermal or adiabatic gas, for the radiation-gas mixture this requires inverting quartics in temperature. This can be achieved numerically, or analytically (Fathi, Mobadersany & Fathi 2012), though these methods prove to be ineffectively slow given that multiple inversions are required per cell per timestep. We instead implement a lookup table (LUT) of pre-calculated thermodynamic values that ATHENA++ then interpolates over. As lookup time does not scale with table size, we are able to load large LUTs (here, $500 \times 500$) into memory; under this treatment calculating thermal quantities can be performed with approximately 6–8 decimal places of precision without major slowdown. We do not expect these minor deviations from the analytical values to have significant effect on the hydro evolution.

### 2.1.3 Heating and cooling

Viscous dissipation is already included in the energy conserving equation (10), heating the gas in shear. The other important contributor to heating is shock heating during both minidisc formation and collision.

Heating is counteracted by radiative cooling in the disc, quantified by $\mathscr{C}$, the cooling rate per unit area

$$\mathscr{C} = 2\sigma T_{\rm eff}^4 = \frac{a_r c T_{\rm eff}^4}{2}, \tag{18}$$

where $\sigma = \frac{a_r c}{4}$ is the Stephan–Boltzmann constant, where $c$ is the speed of light. We rework this into a cooling rate per unit volume $Q$, to fit our equations of fluid motion which use an energy density,

$$Q = \frac{a_r c T_{\rm eff}^4}{4H}. \tag{19}$$

Here, $H$ is the disc scale height and $T_{\rm eff}$ is the effective disc temperature, related to the mid-plane temperature $T$ by the effective optical depth $\tau_{\rm eff}$ (Hubeny 1990)

$$T^4 = \tau_{\rm eff} T_{\rm eff}^4. \tag{20}$$

The effective optical depth is obtained from the optical depth $\tau$

$$\tau_{\rm eff} = \frac{3\tau}{8} + \frac{\sqrt{3}}{4} + \frac{1}{4\tau}, \tag{21}$$

where

$$\tau = \kappa \rho H \,. \tag{22}$$

For simplicity, we assume an optically thick disc $\tau \gg 1$ with opacity $\kappa$ dominated by electron scattering $\kappa = 0.4\,{\rm cm^2\,g^{-1}}$, for which

$$Q = \frac{2a_r c T^4}{3\kappa \rho H^2}. \tag{23}$$

In addition to the radiation + gas simulations mixture, we run isothermal simulations for comparison where we do not follow heating and cooling but assume instantaneous thermal restoration to a fixed temperature.

**Table 1.** Gas properties for the global AGN disc. Defining each of the quantities, $M_{\rm SMBH}$ is the central SMBH mass, $R$ is the radial distance separating the SMBH and the shearing frame centre, $\kappa$ is the gas opacity (set by electron scattering), $\alpha$ is the viscosity coefficient, $l_{\rm E}$ is the ratio of the disc luminosity to the Eddington limit, and $\epsilon$ is the radiative efficiency of accretion.

| $M_{\rm SMBH}$ | $R$ | $\kappa$ | $\alpha$ | $l_{\rm E}$ | $\epsilon$ |
| --- | --- | --- | --- | --- | --- |
| $4 \times 10^6\,{\rm M}_\odot$ | 0.0075 pc | 0.4 cm$^2$g$^{-1}$ | 0.1 | 0.05 | 0.1 |

**Table 2.** Gas properties local to, and homogeneous across, the ambient simulation domain. Defining each of these initial quantities, $\rho_0$ is the gas density, $\Sigma_0$ is the gas surface density, $T_0$ is the temperature (for both radiation and gas), $c_{s,0}$ is the isothermal sound speed, $\beta_0$ is the gas pressure fraction, and $\frac{H}{R}$ is the ratio of disc scale height to distance from the SMBH.

| $\rho_0$ | $\Sigma_0$ | $T_0$ | $c_{s,0}$ | $\beta_0$ | $\frac{H}{R}$ |
| --- | --- | --- | --- | --- | --- |
| $2.65 \times 10^{-11}$ g cm$^{-3}$ | $5.15 \times 10^3$ g cm$^{-2}$ | 2448 K | 6260 m s$^{-1}$ | 0.992 | 0.0042 |

*2.1.4 Viscous prescription*

In the ambient disc, we adopt the local kinematic viscosity of the Shakura–Sunyaev alpha prescription

$$\nu = \alpha c_s H, \qquad (24)$$

where $\alpha$ is a dimensionless coefficient and $c_s = \sqrt{\frac{k_B}{\mu_p m_u}T}$ is the isothermal sound speed. These quantities are readily available within the undisturbed disc by assuming isothermal vertical hydrostatic equilibrium such that $H = c_s/\Omega$ and a given disc opening ratio $H/R$ (where $c_s$ is set by thermal equilibrium between viscous heating and radiative cooling). The inclusion of a BH within the disc complicates this viscosity prescription significantly. The gravity well introduced by the BH will suppress the local disc thickness, decreasing the vertical length scale available for turbulent vortices and so reducing the kinematic viscosity. The formation of a hot minidisc due to an increased viscosity for a larger velocity shear, and the additional heating caused by shocks and tidal forcing will act in the other direction, increasing the local temperature and puffing up the disc. Exactly how turbulence and viscosity should develop in this dynamic system remains an open question. Previous studies of barotropic flows around pre-formed embedded binaries have utilized an adaptive version of equation (24) that consider contributions from the BH to the SMBH potential (Li & Lai 2024, equation 6). While this may be reasonable for the comparatively stable case of a preexisting binary embedded in gas, the gas temperature can rise more rapidly and severely during binary formation in the highly energetic minidisc collisions which may increase the local viscosity well above the ambient level. It is not clear that turbulence should be able to grow on such short time-scales, or how the presence of large-scale shocks (see Section 4.2.2) should interact with this turbulence (and subsequently influence viscosity). The evolution of viscosity in these systems warrants further study, ideally within a 3D context so that the vertical extent of each minidisc can be analysed without the need for assumptions of hydrostatic equilibrium. As such, we opt to keep viscosity within the simulation domain homogeneous and fixed to the ambient value.

## 3 INITIAL CONDITIONS

We simulate our BH encounters within a global AGN disc described by a steady Shakura–Sunyaev alpha disc prescription (Shakura & Sunyaev 1973). In our previous study HW1, local disc density was treated as a free parameter, here the requirement for thermal stability requires the use of a more standard disc model which we adopt from Goodman & Tan (2004). In this model, the local disc parameters can be derived from global properties of the AGN disc. We repeat the equations here, assuming viscosity scales with total pressure.

$$\frac{\beta^{\frac{2}{5}}}{1-\beta} \simeq 0.311 \alpha_{0.3}^{-\frac{1}{10}} \hat{\kappa}^{-\frac{9}{10}} \mu_p^{-\frac{2}{5}} \left(\frac{\epsilon_{0.1}}{l_{\rm E}}\right)^{\frac{4}{5}} M_8^{-\frac{1}{10}} r_3^{\frac{21}{20}}, \qquad (25)$$

$$T \simeq 5.27 \times 10^4 \left(\frac{l_{\rm E}^2 \hat{\kappa} \mu_p \beta}{\epsilon_{0.1} \alpha_{0.3}}\right)^{\frac{1}{5}} M_8^{-\frac{1}{5}} r_3^{-\frac{9}{10}}\, {\rm K}, \qquad (26)$$

$$\Sigma \simeq 2.56 \times 10^5 \left(\frac{\alpha_{0.3}}{\beta}\right)^{-\frac{4}{5}} l_{\rm E}^{\frac{3}{5}} \epsilon_{0.1}^{-\frac{3}{5}} \hat{\kappa}^{-\frac{1}{5}} \mu_p^{\frac{4}{5}} M_8^{\frac{1}{5}} r_3^{-\frac{3}{5}}\, {\rm g\,cm}^{-2}. \qquad (27)$$

The newly introduced symbol definitions are listed below: 0.4 cm$^2$ g$^{-1}$ $\hat{\kappa}$ is the gas opacity, $0.3\alpha_{0.3}$ is the alpha disc viscosity parameter, $0.1\epsilon_{0.1}$ is the radiative efficiency of the disc, $l_{\rm E}$ is the ratio of the disc luminosity to the Eddington limit, $10^8 M_8 M_\odot = M_{\rm SMBH}$ and $10^3 R_s r_3 = R$, where $R_s = \frac{2GM}{c^2}$ is the Schwarzschild radius. We list our global AGN properties in Table 1. We initialize our shearing box as a patch of this AGN disc, drawing gas properties from the global disc structure and homogenizing them across the simulation domain,

This ambient system is gas dominated; under the assumption of isothermal vertical hydrostatic equilibrium we predict the local scale height $H = \frac{c_s}{\Omega}$, which in turn sets the kinematic viscosity $\nu = \alpha c_s H$. These local properties are recorded in Table 2. Our shearing box is positioned at $R = 0.0075$ pc, motivated by the semi-analytical studies (Tagawa et al. 2020a), which reported frequent binary mergers near this radius. We expect binary formation to behave differently at different positions within the disc, and also for discs with different global parameters e.g. different accretion rates or $\alpha$. We leave the study of binary formation as a function of disc position and global disc properties for future work.

The resulting patch of AGN disc is gas pressure dominated ($\beta \sim 1$), and stable against gas self-gravity ($Q > 1$; Toomre 1964). Within the simulation, we inject two stellar-mass BHs with $m_{\rm BH} = 25\,{\rm M}_\odot$. We limit our study to a single choice of equal mass interactors, though in principle we expect binary formation to behave differently for interactors of unequal mass. For this BH-SMBH mass ratio, we derive single and binary Hill radii of $r_{\rm H,s} = 10^{-4}$ pc and $r_{\rm H} = 1.2 \times 10^{-4}$ pc. The disc is more dense than those studied in HW1, which considered a range of ambient disc densities with $m_{\rm H,0} = [1.3 \times 10^{-4}, 1.3 \times 10^{-2}] m_{\rm BH}$, where $m_{\rm H,0}$ is the ambient Hill mass: the total gas mass within $r_{\rm H}$ of each BH. In this study, $m_{\rm H,0} \simeq 0.05 m_{\rm BH}$ though by the time the BHs reach close encounter their minidiscs grow considerably denser such that $m_{\rm H} \simeq 0.2 m_{\rm BH}$.

Here, the effect of gas self-gravity may become significant. Each BH is given sufficient time to grow a morphologically stable minidisc, but in principle, evolving the system for longer can allow for more massive minidiscs; we are limited here by the size of the shearing box. Due to computational expense, gas self-gravity is neglected from this work, but merits caution in future studies, especially for higher ambient densities.

A detailed description of the computation domain can be found in HW1, we utilize the same base resolution and number of adaptive mesh refinement levels. We use a larger BH-gas smoothing length $h = 0.05 r_{H,s}$, which was found to improve thermal stability at early times (see Appendix A for a discussion of this choice). The BHs are launched on Keplerian trajectories, such that their relative velocity is a function of their initial radial separation $b$ that is varied between simulations. In order to preserve a similar flight time to encounter between simulations, the initial BH azimuthal separation $\Delta\phi$ is a function of $b$, normalized to a standard radial separation $b_0 = 2.5 r_H$ such that $\Delta\phi(b_0) = 22°$ (a procedure developed in CR2)

$$\Delta\phi(b) = \frac{b}{b_0} \times 22°. \tag{28}$$

These varying azimuthal separations correspond to different linear separations in the $y$-direction of the shearing frame. Varying the azimuthal separation in this way sets the approximate time to encounter to be 35–40 yr. This normalization allows for better comparison between impact parameters, as each system has a roughly equivalent time to accumulate minidisc mass before close encounter.

## 4 RESULTS

We discuss a suite of simulations within a single AGN disc prescription (as detailed in Section 3), but vary the BH initial conditions and fluid equation of state. We vary the initial radial separations of the BHs across $b \in [1.7, 2.6] r_H$ with 19 linear spaces and consider the fluid to evolve either as an isothermal pure gas or as an adiabatic mixture of radiation and gas. This ensemble allows us to scrutinize the effects of the EoS on gas-assisted BBH capture, but also reveals the new phenomenon of periodic disc novae, as we will describe in this Section. We do not consider interactions within different AGN discs, or at different positions within said discs. While the effects of such variations certainly warrants study, we leave this for future investigations. After a brief overview of the definitions the results are split into two sections each considering a key question. In Section 4.2, how does changing the equation of state affect the minidisc structure and hydrodynamic evolution during BH close encounters? In Section 4.3, how does changing the equation of state influence the likelihood of successfully BBH formation?

### 4.1 Definitions

In discussing the evolution and outcome of these simulations, we use much of the language implemented in HW1. We consider a total binary mass $M_{bin} = m_1 + m_2$ and reduced binary mass $\mu = m_1 m_2 / M_{bin}$, along with binary angular momentum $L_{bin}$, centre-of-mass energy $E_{bin}$ and eccentricity $e$,

$$L_{bin} = \mu |(\mathbf{r_1} - \mathbf{r_2}) \times (\mathbf{v_1} - \mathbf{v_2})|, \tag{29}$$

$$E_{bin} = \frac{1}{2}\mu|\mathbf{v_1} - \mathbf{v_2}|^2 - \frac{GM_{bin}\mu}{|\mathbf{r_1} - \mathbf{r_2}|}, \tag{30}$$

$$e = \sqrt{1 + \frac{2E_{bin}L_{bin}^2}{G^2 M_{bin}^2 \mu^3}}, \tag{31}$$

here, $\mathbf{r_i}$ and $\mathbf{v_i}$ are the positions and velocities of BH $i$ and $G$ is the gravitational constant. For systems with $E_{bin} < 0$ (bound in the conventional, isolated sense), we can define a semimajor axis $a$

$$E_{bin} = -\frac{GM_{bin}\mu}{2a} \tag{32}$$

allowing us to further define a standard energy scale $E_H$, the Hill energy. This is the absolute binary energy of a system with semimajor axis $a = r_H$, such that $E_H = \frac{GM_{bin}\mu}{2r_H}$.

### 4.2 Morphology and encounter chronology

Here, we discuss the hydrodynamic distinctions between the minidisc structures arising from each equation of state and analyse the evolution of the binary system as the components pass through close encounter. We calculate the luminosities and spectra of the simulations as they develop.

#### 4.2.1 Pre-encounter minidiscs

We consider the structure of the minidiscs before BH–BH close encounter occurs. Fig. 2 depicts spatial maps of standard minidiscs from an isothermal and radiative adiabatic system at around $t = 30$ yr (here, the BHs are $\sim 2r_H$ apart). We measure $m_H$ and $\mathcal{L}_H$ as the total gas mass and luminosity within the Hill sphere $r = r_H$ (the Hill mass and Hill luminosity, respectively). We further define $R_m$ and $R_\mathcal{L}$ as the bounding radii within which 75 per cent of the Hill mass or Hill luminosity can be found. These statistics are recorded in Table 3. We see that the Hill masses are roughly identical for the isothermal gas runs and runs with gas + radiation mixture, with $m_H \sim 0.2 m_{BH}$. The substructures within each minidisc system are very different however. The isothermal system is much denser, with thinner spiral streamers and a less extended disc: using the $R_m$ metric the radiative disc is ~60 per cent larger. This is not unexpected, the isothermal system is cooler and so lacks the pressure support to maintain the larger, diffuse minidiscs of the radiative EoS. The radiative minidisc covers a wide range of thermal regimes, from the centre of the streamers which are strongly gas dominated ($\beta \sim 0$), to the streamer edges that are shock heated to an intermediate regime where gas and radiation support equally ($\beta \sim 0.5$), to the hot minidisc core where radiation dominates ($\beta \sim 1$). The radiative system lacks the relatively well defined spiral arms of the isothermal minidisc and features an extended turbulent core.

We are unable to estimate the luminosity of the isothermal system due to the assumptions implicit to its hydrodynamic evolution (namely instantaneous cooling), but we include the luminosity profile for the radiative system. The radiative minidisc is much hotter and brighter than the ambient gas, with $\mathcal{L}_H \sim 2\mathcal{L}_0$ where $\mathcal{L}_0$ is the total emission of the ambient domain, equivalent to $5 \times 10^{38}$ erg s$^{-1}$. Alternatively, $\mathcal{L}_H \sim 0.3 L_{edd,BH}$. Care should be taken when considering hydrodynamic data near the minidisc centre. While the simulations presented are high resolution and feature a reasonably small smoothing length $h$, the inability to resolve the inner accretion disc down to its inner edge is a non-convergent issue. For comparison, $h = 0.05 r_{H,s}$, equivalent to $10^6 r_s$ where $r_s$ is the Schwarzschild radius of the BHs. We do not expect such sub-resolution physics to have a major effect on the binary dynamics, but resolving the inner minidisc should result in an increase in minidisc luminosity due to the higher temperatures predicted there. Thus, we include these minidiscs statistics here, but with an understanding that these are the best recoverable quantities rather than a truly accurate depiction of the small-scale system.[1]

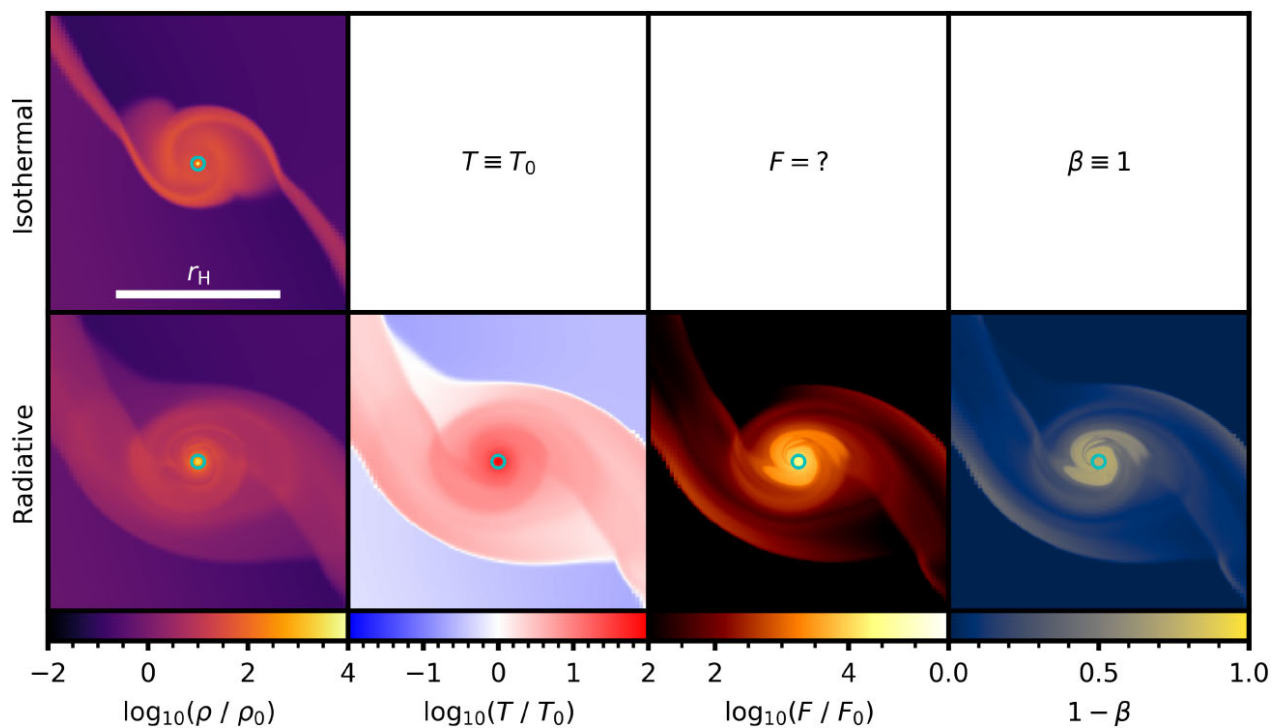

**Figure 2.** Spatial maps of a variety of hydrodynamic quantities for standard pre-encounter minidiscs featuring both equations of state. Here, $\rho$, $T$, $F$, and $\beta$ are the gas density, fluid temperature, thermal emission, and gas pressure fraction. The cyan ring marks the smoothing radius $h = 0.05 r_{H,s}$ about the BH. We note clear morphological differences between the minidiscs under different thermal prescriptions: the radiative disc is larger, more diffuse, and considerably hotter. The central minidisc is hot enough to feature substantial contributions from radiation, emphasizing the importance of including radiation in the equation of state. Data are not plotted for those quantities fixed by the equation-of-state assumptions.

**Table 3.** Total gas mass and luminosity integrated over the Hill sphere for minidiscs of varying equations of state at $t = 30$ yr, when the BHs are approximately $2r_H$ apart. $R_m$ and $R_\mathcal{L}$ are the radii within which 75 per cent of the Hill mass ($m_H$) and Hill luminosity ($\mathcal{L}_H$) can be found. The principal difference between discs of varying EoS is that the radiative minidiscs are considerably more diffuse, exemplified by the larger $R_m$ value for the same $m_H$, see Fig. 2. Gas masses are compared to the single BH mass $m_{BH} = 25\,M_\odot$, luminosities to the total output of the ambient simulation $\mathcal{L}_0 = 5 \times 10^{38}$ erg s$^{-1}$.

| EoS | Isothermal | Radiative |
|---|---|---|
| $m_H [m_{BH}]$ | 0.21 | 0.185 |
| $\mathcal{L}_H [\mathcal{L}_0]$ | N/A | 2.13 |
| $R_m [r_H]$ | 0.28 | 0.45 |
| $R_\mathcal{L} [r_H]$ | N/A | 0.23 |

### 4.2.2 Disc novae

The most significant discrepancy between the behaviour of models with varying equations of state can be found during and after the first close encounter between the BHs. In the isothermal systems of HW1, CR1 and CR2, disc collisions featured mass stripping from the minidiscs in the form of long streamers, with gas mass torn kinetically from around the BHs (see HW1, Fig. 4). This system is modified once the assumption of instantaneous cooling is relaxed; when two BHs undergo severe close encounter and their inner minidiscs collide a massive central overpressure is generated, driving a strong blast wave which propagates over many Hill radii. These explosions are 'disc novae', quasi-circular blast waves that propagate radially outwards under the support of a central thermal energy ejection. The shock front advance is analogous to the Sedov–Taylor blast wave solution (Taylor 1950; Sedov 1959) with some complications, primarily due to the inhomogeneous background that the shock must propagate through in this case. Furthermore, the energy injection, while strongly localized to the periapsis, is not truly instantaneous, though the shock heating period is brief enough to drive a similar morphology.

The general chronology of a close encounter with blast wave production is depicted in Fig. 3. When the two BH minidiscs collide around $t = 35$ yr, the shocking of minidisc gas drives a strong overpressure which forces a blast wave outwards. The blast is quite asymmetric, the shock propagates fastest in the $y$ direction, perpendicular to the radial direction to the SMBH, as there are no spiral overdensities in this direction. The shock continues to expand, sweeping mass from the mutual Hill sphere. At around $t = 36$ yr, the young binary passes through periapsis again, generating a new shock wave. Each periapsis features the generation of a new shock wave: these shocks tend to be more circular than the original as the background through which the shock propagates is now more homogeneous. As the binary hardens impulsively by gas gravitation during each periapsis passage, the binary period shortens and the frequency of shock generation increases. The strength of the blasts tend to decrease with time as more mass is ejected from the system.

---

[1] Note that for a Shakura–Sunyaev disc the flux scales with $r^{-3}$ and the total luminosity may be dominated by the unresolved minidisc possibly up to $\mathcal{L}_{tot} \sim d/r_s \mathcal{L}_{sym}$, where $\mathcal{L}_{sym}$ is the luminosity obtained with an inner boundary of $d$, although the actual luminosity may be limited by radiation pressure to around $L_{edd,BH}$.

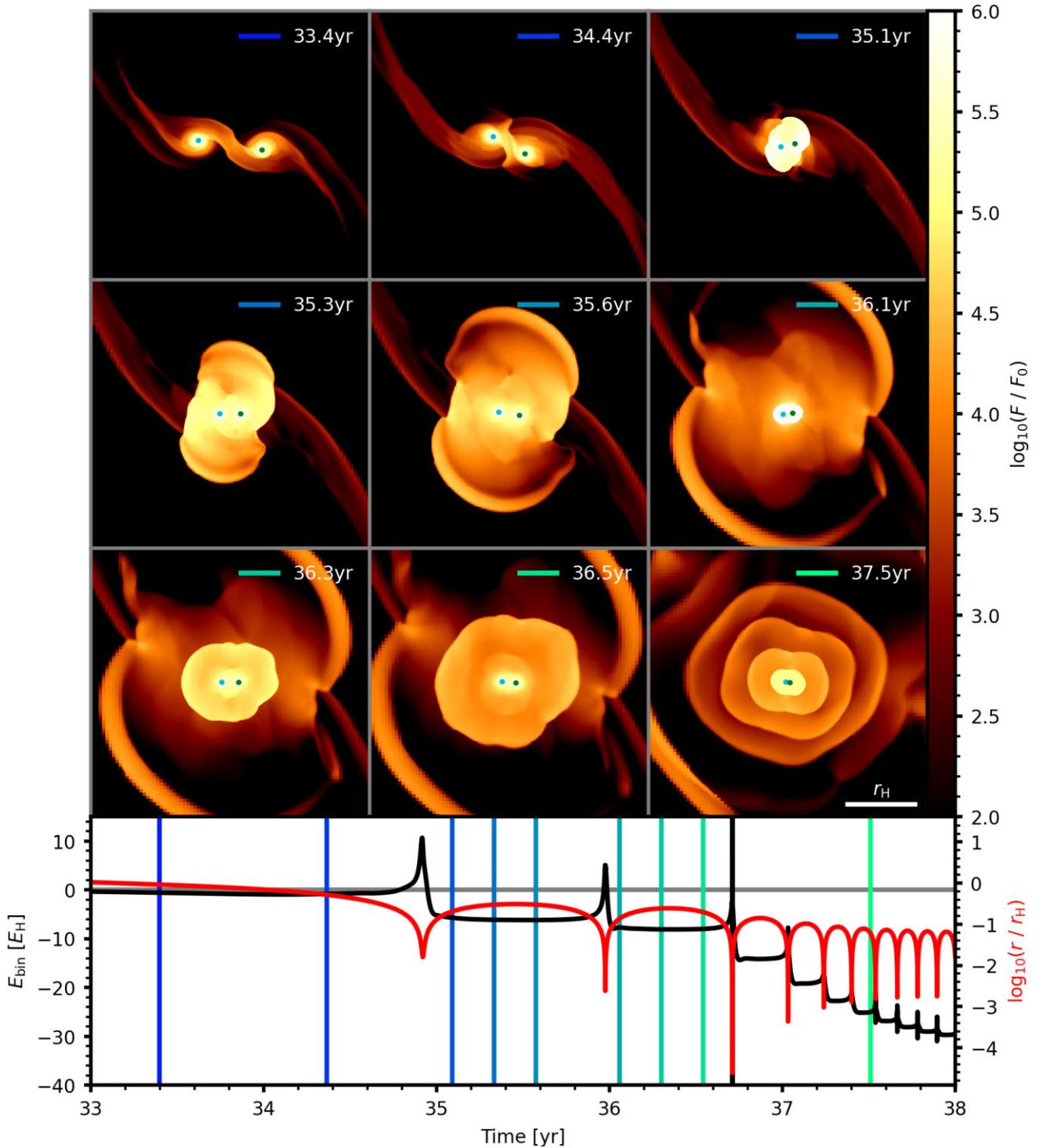

**Figure 3.** Spatial flux density maps for the radiative model with $b = 2.2r_H$, featuring strong shock generation during the first close encounter. Here, $F$ is the thermal emission per unit area as given by the cooling rate (see equation 18). The cyan and green circles represent the smoothing radii $h = 0.05r_{H,s}$ about each BH. The lower panel depicts the evolution of the binary energy $E_{bin}$ and BH separation $r$, along with the times of the flux snapshots. In the first row, the BHs approach each other, generating bright shocks as the edges of their minidiscs collide. At $t = 34.9$yr the binary passes through periapsis, generating a strong central overpressure as the high density/temperature minidisc cores collide at high speed. The blast wave generated by this pressure propagates radially outwards (shown in the middle row), generating a luminous, elliptical fireball. Dissipation during the first encounter is sufficient to form a binary; as the binary undergoes further periapsis passages it generates more blast waves (shown in the lower panels). As the binary continues to harden, its period decreases, resulting in blasts of increasing frequency. Hardening efficiency decreases with time as the blast waves deplete the Hill sphere of gas mass, generating a hot underdense region around the binary. While the rise and fall of energy during each encounter is real, the extreme energy peak during the very deep third periapsis at $t = 36.7$ yr is due to round-off errors, addressed in Appendix B. See the solid orange line in Fig. C1 for the BH trajectories.

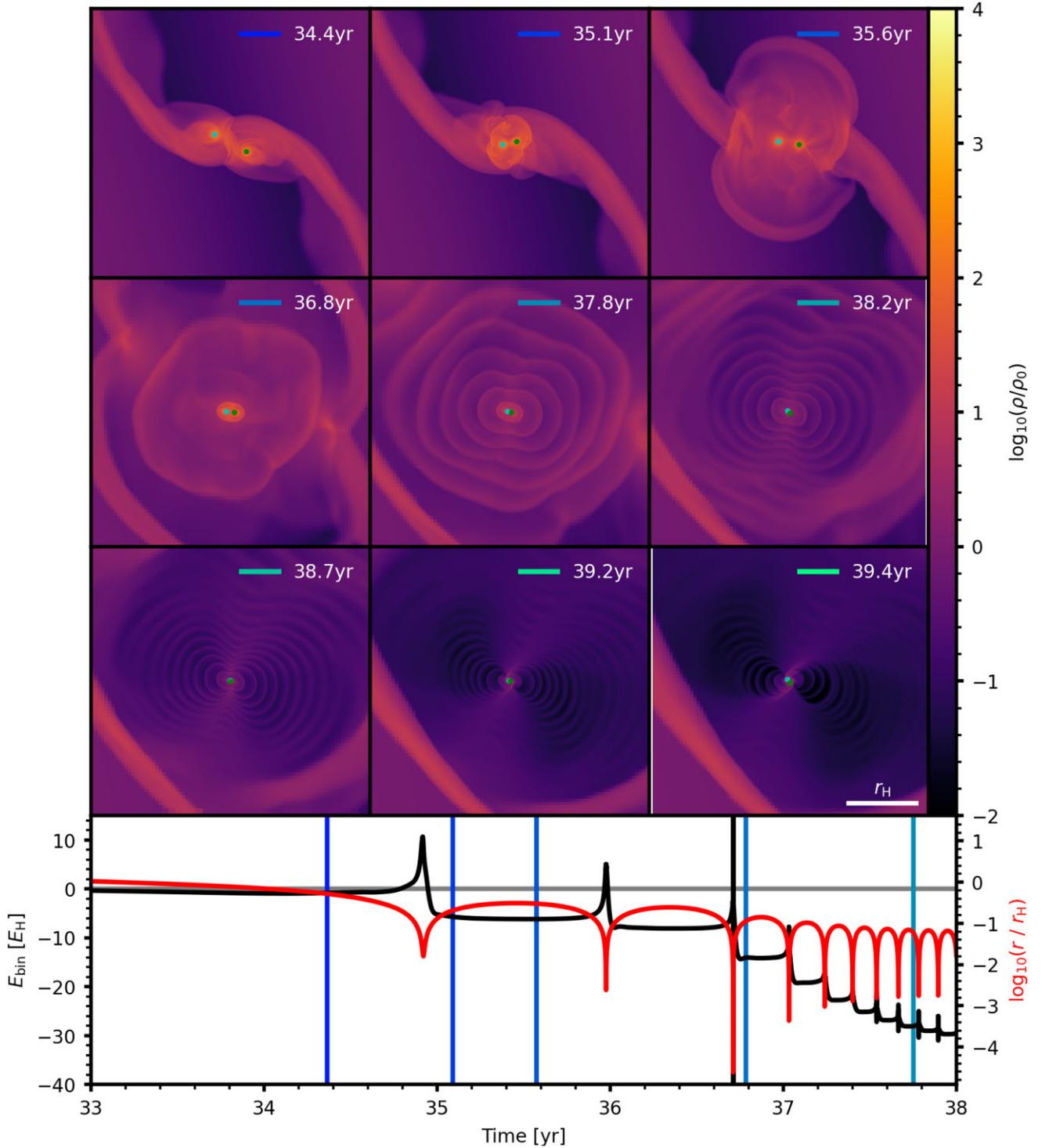

**Figure 4.** Spatial gas density maps for the radiative model with $b = 2.2 r_H$ (as in Fig. 3), showing both the initial explosion and late time evolution. The cyan and green circles represent the smoothing radii $h = 0.05 r_{H,s}$ about each BH. Each of the shock fronts are visible as edge-dense rings in the gas density. Once the shock strength has fallen off, the binary is left in a hot, underdense vacuum. Shocks are still generated during each periapsis, but the intensity has dropped significantly as there is less mass left in the minidiscs to collide with. See the solid orange line in Fig. C1 for the BH trajectories.

The continual injection of thermal energy into the circumbinary environment generates a hot, underdense region which prevents external AGN disc mass from joining the system. Fig. 4 depicts the same system as Fig. 3, but this time shading by gas density at late times. We see that at later times, the binary hardening has slowed, the outbursts are weaker and the binary is surrounded by hot, low density gas. The binary centre-of-mass initially rests in a position of unstable equilibrium and at late times asymmetric gas gravitation tends to knock the binary from its original position: from here inertial forces will perturb it further away. As such the binary will eventually encounter undisturbed, colder gas in regions distant to the original formation location. We focus primarily on the encounter, and early post-encounter evolution in this study. Once a binary reaches $E_{\rm bin} = -20E_{\rm H}$, the binary semimajor axis is the same size as the gravitational smoothing length $d$. The use of a finite, fixed smoothing length for gas gravitation means that at late times when the binary is very hard, the circumsingle/circumbinary gas flows are unlikely to be faithfully represented.

The generation of large-scale blast waves during minidisc close encounters is novel to the radiative minidisc-collision system, as yet underexplored in the literature. Similar outburst systems have been considered in AGNs, such as embedded supernovae (Grishin et al. 2021) and star–disc collisions ((Jane) Dai, Fuerst & Blandford 2010; Tagawa & Haiman 2023). We note that while the generation of an outflow is likely to be a natural consequence of realistic minidisc collisions, the exact form of this outflow is unlikely to be consistent between this 2D study and a full 3D simulation. We do not expect the blast wave to remain confined without the plane of the AGN disc; indeed it seems likely that the blast should preferentially escape vertically from the disc as the path of least resistance, as observed in Grishin et al. (2021) (figs 9 and 10). This could lead to the generation of wind-like phenomena with substantial high-temperature gas mass being ejected out of the disc plane.

Nonetheless, while it seems likely that a lot of the 3D physics pertinent to these collisions will be misrepresented within a 2D simulation, we still observe phenomena that we expect to be generic to the disc novae scenario. Irrespective of dimensionality, high-velocity collisions between gas minidiscs during deep close encounters will drive strong shock heating. This heating will be driven back into the surrounding gas, and while the form of this feedback is potentially ambiguous, this injection of heat is likely to have a profound effect on the binary/circumbinary environment. Perhaps most importantly, such shock heating is expected to result in significant thermal emission, discussed in greater detail in Section 4.2.3.

The existence of blast waves within the simulation complicates predictions as to local viscosities, and is the principle reason why we do not attempt to implement an adaptive viscosity prescription. It is not clear how such strong shocks should interact with turbulence in the steady disc, making predictions of viscosity impossible. This system clearly warrants further study in 3D, where estimations of vertical scales may allow for better descriptions of turbulence. Such studies would also provide a clearer picture of how the thermal energy generated in the disc collisions are fed back into the local AGN disc.

### 4.2.3 Outburst luminosity

Optically thick cooling included explicitly in the code allows us to analyse the thermal emission of the minidisc systems and present potentially observable phenomena. In discussing the luminosity of these systems, we can compare both to the ambient local emission $\mathcal{L}_0$ and the total AGN emission $\mathcal{L}_{\rm AGN}$. The former can be calculated from the homogeneous, ambient disc conditions: integrating over the simulation surface area $A$ produces $\mathcal{L}_0 = \mathscr{C}A \sim 5 \times 10^{38}\,{\rm erg\,s^{-1}}$.

Our disc model is parametrized with respect to its Eddington luminosity $\mathcal{L}_{\rm E}$ such that $\mathcal{L}_{\rm AGN} = l_{\rm E}\mathcal{L}_{\rm E}$. For this study, using the data from Table 1, we derive

$$\mathcal{L}_E = \frac{4\pi G M_{\rm SMBH} m_p c}{\sigma_{\rm T}} \simeq 5 \times 10^{44}\,{\rm erg\,s^{-1}}, \qquad (33)$$

$$\mathcal{L}_{\rm AGN} \simeq 2.5 \times 10^{43}\,{\rm erg\,s^{-1}}. \qquad (34)$$

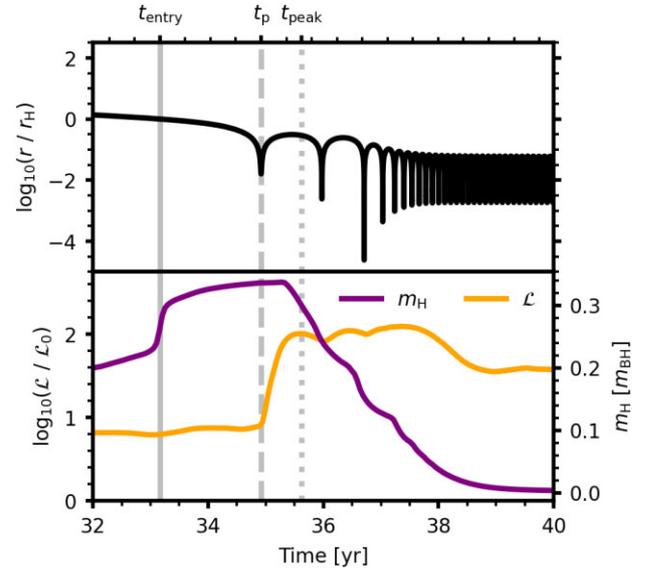

**Figure 5.** Luminosity evolution for a radiative model with $b = 2.2r_{\rm H}$, showing the strong increase in thermal emission ($\times 10$) immediately after first periapsis due to the fireball generated by severe minidisc collision. The upper panel depicts the binary separation $r$ and the lower the luminosity and Hill mass. Three times are marked by vertical lines: when the BHs first enter their mutual Hill sphere, when the BHs pass through first periapsis and when the luminosity first peaks ($t_{\rm entry}$, $t_{\rm p}$, and $t_{\rm peak}$, respectively). The luminosity rises rapidly over a period of around $t_{\rm peak} - t_{\rm p} = 0.7$ yr, after which it begins to fall. Subsequent periapsis passages are associated with further bumps in luminosity, though once the binary period shortens sufficiently these peaks become impossible to distinguish. The blast waves generated during outbursts reduces $m_{\rm H}$, the mass in the Hill sphere. The luminosity begins to decrease consistently around 37.5 yr, likely due to the weakening outburst strength.

We see that the ambient luminosity of the simulation represents only a tiny contribution to the total AGN luminosity, around five orders of magnitude beneath the total output.

Each BH, in drawing gas mass into a minidisc, acts as a source of heat in the simulation, generating shocks and shears that thermalize gas kinetic energy. As such, even pre-close-encounter the embedded systems is around 10 times brighter than the ambient shearing box. Fig. 5 depicts the luminosity evolution of the radiative system with $b = 2.2r_{\rm H}$, as well as the Hill mass. There are three times of interest, $t_{\rm entry}$ when the BHs first penetrate the mutual Hill sphere, $t_{\rm p}$ when the binary passes through first periapsis and $t_{\rm peak}$ when luminosity peaks in the first outburst. At $t_{\rm entry}$ there is an increase in $m_{\rm H}$ due to the minidiscs beginning to coalesce. At $t_{\rm p}$ the rapid collision of the inner minidiscs generates strong shock heating and an expanding fireball. The luminosity of the system rises as the fireball expands, reaching a local maximum at $t_{\rm peak}$. At this point, the shearing box is 100 times brighter than its ambient state. After the first peak, there are further fluctuations in luminosity associated with subsequent outbursts generated in later periapsis passages; these fluctuations are relatively minor. After a few orbits these fluctuations occur rapidly enough to be indistinguishable from each other, and over time the luminosity begins to fade as the outbursts weaken in intensity. This is perhaps due to the depletion of gas mass from the Hill sphere, resulting in weaker shock generation during circumsingle disc collisions.

As the luminosity of the system rises, the Hill mass $m_{\rm H}$ decreases. Mass is blown out of the Hill sphere by the strong blast waves, resulting in periodic dips in $m_{\rm H}$ as the binary goes through periapsis.

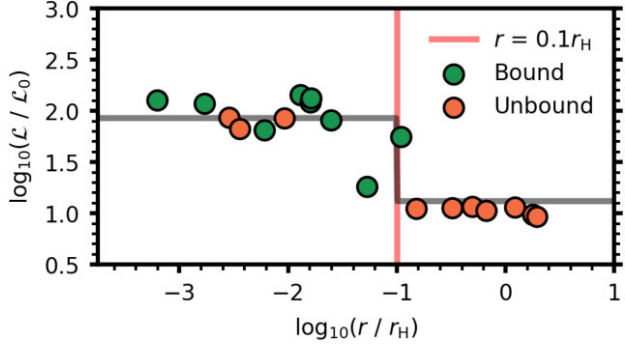

**Figure 6.** Peak luminosity for all radiative models. These peak luminosities are achieved immediately after the first close encounter, with emission dominated by the hot fireball generated by the minidisc collisions. The $\mathcal{L}$–$r_p$ relation is well approximated by a step function at $r = 0.1 r_\mathrm{H}$, equivalent to the size of the hotter inner minidisc. For encounters with $r < 0.1 r_\mathrm{H}$, deeper encounters are not associated with greater peak luminosities. Bound systems are those that achieve $E_\mathrm{bin} < -2 E_\mathrm{H}$ and so are considered to be stable against SMBH ionization, all other systems are unbound and fail to form stable binaries.

After only a few periods, $m_\mathrm{H}$ has dropped beneath ambient levels and the binary is left within a hot underdense pocket. This is not an equilibrium state: the rate of heat generation by the periapsis collisions decreases as $m_\mathrm{H}$ falls and eventually the vacuum will cool sufficiently to allow gas to flow back on to the binary system. The binary, existing at a point of unstable equilibrium within the shearing box, tends to be ejected from the box before this cooling phase is reached, encountering new gas mass as it travels through more distant, undisturbed regions of the disc.

While this luminosity chronology is very similar across strongly colliding systems, not all flyby configurations undergo severe minidisc collisions. Fig. 6 records the peak luminosities achieved in the radiative simulations and compares it to the depth of first periapsis. We note a step-up in peak luminosity for systems with a suitably deep first encounter, of around an order of magnitude. The threshold depth for outbursts is around $0.1 r_\mathrm{H}$, comparable to the size of the hot, radiation dominated minidisc core. The size of this core may depend on the smoothing length $d = 0.05 r_\mathrm{H}$, as for $r < d$ the local field strength diverges from the Newtonian solution, or on the resolution. Better resolving the flow near the BHs may change the degree of shock/viscous heating in pre-collision minidiscs, and so affect the inner minidisc properties. Systems with collisions deeper than the $r \sim 0.1 r_\mathrm{H}$ threshold feature strong minidisc core collisions, generating high temperatures via shock heating. Without a suitably violent close encounter, the peak luminosity is similar to the sum of the two isolated minidisc luminosities. Strong luminous outbursts are well correlated with binary formation, as both require deep first close encounters. Of the 12 models exhibiting major rapid increases in luminosity, 9 result in successful binary formation.

We find the post-periapsis luminosity peaks around $10^2 \mathcal{L}_0$, equivalent to around $2 \times 10^{-3} \mathcal{L}_\mathrm{AGN}$, with a rise time of around 0.5–1 yr (see Fig. 5). These outbursts are likely to be subdominant to other sources of variation within the AGN disc; applying the variability modelling of Kelly, Bechtold & Siemiginowska (2009) to our system, we might expect $R$-band variability of a similar magnitude over a period of only a month. These events may be more prominent within specific spectroscopic bands that we address in Section 4.2.4.

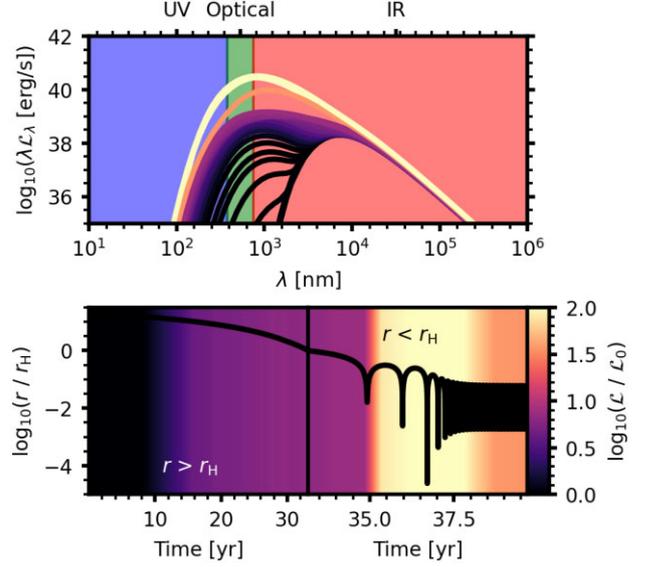

**Figure 7.** Rest-frame spectral power evolution for the radiative system with $b = 2.2 r_\mathrm{H}$. The top panel includes spectra for 30 evenly spaced intervals in time, coloured by the luminosity for that snapshot (see colourbar in bottom panel). In the bottom panel we overplot the BH separation on top of the total luminosity, both as a function of time. The moment the BHs enter their mutual Hill sphere is indicated by the vertical line at about 31.5 yr. The domain spectrum after the BHs have formed their minidiscs (by $t \sim 20$ yr) is much hotter and bluer than ambient, with the minidisc component dominating the spectrum. After the first close encounter the luminosity spikes, with the emission peaking in the optical and near IR. Once the outburst intensity dies down, the spectrum cools somewhat, but remains substantially brighter than ambient.

### 4.2.4 Outburst spectra and prominence

As presented in Section 4.2.3, the post-periapsis disc novae, while 100 times brighter than the local ambient disc, are too dim to be discerned from other sources of variation in the bolometric AGN luminosity. We consider whether these outbursts might be identifiable spectroscopically by predicting the spectral energy density associated with the thermal emission. These spectra are generated by integrating a series of Planck functions over the entire simulation domain.

$$\mathcal{L}_\lambda = \frac{2hc^2}{\lambda^5} \int_A \frac{dA}{\exp\left(\frac{hc}{\lambda k_\mathrm{B} T_\mathrm{eff}}\right) - 1} \simeq \frac{2hc^2}{\lambda^5} \sum_{i=1}^{N_c} \frac{A_i}{\exp\left(\frac{hc}{\lambda k_\mathrm{B} T_{\mathrm{eff},i}}\right) - 1}. \tag{35}$$

Here, $\lambda$ is the wavelength, $h$ is Planck's constant, and $k_\mathrm{B}$ is the Boltzmann constant, with $A_i$ and $T_{\mathrm{eff},i}$ the surface area and effective temperature for cell $i$ of $N_c$ cells total. We first note that the spectrum of the system with embedded black holes (before close encounter) is substantially different from the local ambient disc: the spectrum is brighter and significantly bluer, peaking in the rest-frame optical/near-IR (see Fig. 7). At early times as the disc is forming, the spectrum is well described by a large cool component, and a smaller hot component; this is visible in the twin peaks of the early spectrum. After around $t = 20$ yr, however, the spectrum of the simulated region is effectively dominated by the minidiscs at all wavelengths.

When the BHs undergo first close encounter a major blast wave is generated, and the emission peaks in the optical/near IR. In Fig. 8 we compare the outburst spectra to the background AGN emission by considering the effective temperature of a disc heated only by viscous dissipation (see equation 3 from Goodman 2003). While the outbursts are subdominant in the bolometric emission, the AGN spectrum peaks in the UV, compared to the outbursts' peak in the near IR. Limiting the integrated luminosity to the IR bands ($\lambda > 750$nm), the outbursts are significantly more prominent, with $\mathcal{L}_{\text{peak,IR}} = 0.03 - 0.04 \mathcal{L}_{\text{peak,IR}}$. While still subdominant to the AGN IR emission, these outbursts may be able to contribute to variability in the IR bands.

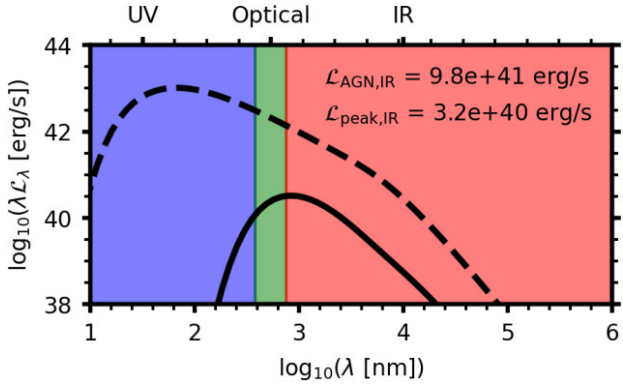

**Figure 8.** Comparison of rest-frame spectral power between the global ambient AGN disc (dashed line) and the $b = 2.2 r_H$ simulation domain at peak luminosity (solid line). While the outbursts are still dimmer than the AGN at all frequencies, they peak at longer wavelengths and so are more prominent in the IR bands than they are bolometrically. Integrating over the IR wavelengths, this outburst peaks at $\mathcal{L}_{\text{peak,IR}} \sim 0.03 \mathcal{L}_{\text{AGN,IR}}$.

We note that the exact morphology and evolution of these outbursts are unlikely to be faithfully captured due to the restricted nature of this study, see Section 5.2 for a discussion of these limitations

### 4.3 Capture likelihood

Here, we consider how variations to the equation of state may affect the likelihood of embedded binary black hole formation. We calculate the orbital energy dissipated from each system during close encounter and compare systems with varying impact parameters.

#### 4.3.1 Dissipation efficiency

The distinguishing characteristic between systems that form binaries and those that do not is the amount of orbital energy that can be dissipated during the first close encounter. This dissipation can be driven by the SMBH or by gas gravitation, with the latter tending to dominate. Each of these quantities can be derived from integrating the dissipation rates $\epsilon_{\text{SMBH}}$ and $\epsilon_{\text{gas}}$, over the close encounter.

$$\epsilon \equiv \frac{d}{dt}\left(\frac{E_{\text{bin}}}{\mu}\right) = \epsilon_{\text{gas}} + \epsilon_{\text{SMBH}}, \quad (36)$$

$$\epsilon_{\text{SMBH}} = (\mathbf{v_1} - \mathbf{v_2}) \cdot (\mathbf{a_{1,\text{SMBH}}} - \mathbf{a_{2,\text{SMBH}}}), \quad (37)$$

$$\epsilon_{\text{gas}} = (\mathbf{v_1} - \mathbf{v_2}) \cdot (\mathbf{a_{1,\text{gas}}} - \mathbf{a_{2,\text{gas}}}), \quad (38)$$

where $\mathbf{v_i}$ and $\mathbf{a_i}$ are the velocity of and acceleration on BH $i$, here acceleration is exerted on the BHs by frame forces (SMBH) or by gas gravitation (gas). We then calculate net dissipation from each source as

$$\Delta E_{\text{SMBH}} = \int_{r=r_H}^{r=r_p} \mu \epsilon_{\text{SMBH}} \, dt, \quad (39)$$

$$\Delta E_{\text{gas}} = \int_{r=r_H}^{r=r_a} \mu \epsilon_{\text{gas}} \, dt, \quad (40)$$

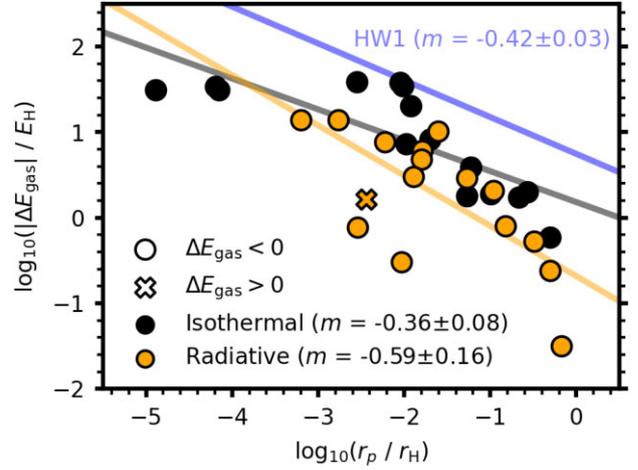

**Figure 9.** Strength of gas dissipation for both isothermal and radiative models, with fittings in the form $\Delta E_{\text{gas}} \propto r_p^m$ where $r_p$ is the depth of first close encounter. The model from HW1 is included in blue. The trends are broadly consistent, with dissipation increasing with depth, however dissipation is generally weaker than the extrapolated model of HW1. We also note a single run in the radiative set which injects energy ($\Delta E_{\text{gas}} > 0$) during the encounter.

where integration is performed from first Hill intersection at $r = r_H$, through to first periapsis $r_p$ or first apoapsis $r_a$. If no binary forms (and hence there is no first apoapsis) $\Delta E_{\text{gas}}$ is integrated until the BHs escape the Hill sphere. The integration limits in these definitions are inherited from the predictive modelling of HW1, as they best decoupled the SMBH and gas gravity effects. The SMBH dissipation is a function of the BH trajectory and post-periapsis the trajectory is strongly dependent on the severity of gas dissipation at periapsis. Halting the SMBH integrating at periapsis prevents cross-interaction between the two effects and allows for the best measure of energy soon after the periapsis ($\epsilon_{\text{gas}}$ is strongest very close to periapsis, whereas $\epsilon_{\text{SMBH}} \to 0$ at periapsis).

HW1 found that the gas dissipation during first close encounter $\Delta E_{\text{gas}}$ was dependent on the minidisc mass (and therefore the ambient gas density $\rho_0$) and the initial periapsis depth $r_p$. Fitting over a series of 345 simulations, HW1 derived $\Delta E_{\text{gas}} \sim r_p^{-0.43 \pm 0.03}$, this was supported by CR2 which found $\Delta E_{\text{gas}} \propto r_p^{-0.42 \pm 0.16}$. Fig. 9 compares the scaling of gas dissipation with periapsis depth for the isothermal and radiative systems and includes fits in the form $\Delta E \propto r_p^m$. We recover the same general relationship, deeper first close encounters tend to dissipate more energy by gas gravitation. The slopes $m$ are within $1\sigma$ of the HW1 fit, though we note the very shallow and very deep encounters appear to deviate from this relationship.

The model from HW1 tends to overestimate the dissipation of the isothermal system. In that work, it was noted that the fit performed best at lower ambient gas densities where the gas did not strongly perturb the BH trajectories pre close encounter: here, the ambient disc density is higher than even the most dense systems analysed in that work. We note that the isothermal models tend to dissipate more energy than their radiative counterparts. We reason this increased efficiency may be due to the compactness of the isothermal systems: an increase in gas mass closer to the BHs during close encounters will result in stronger gravitational attraction (see Table 2). The effect is relatively minor for most of the models however, and we only have a limited number of simulations to infer from.

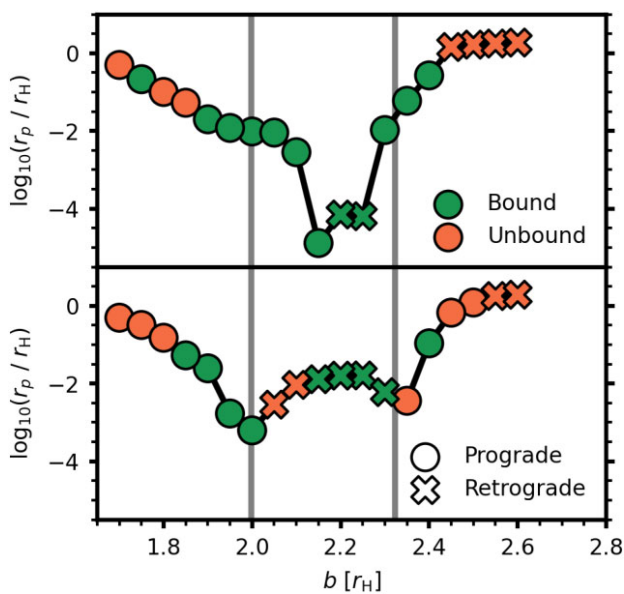

**Figure 10.** Depth of first close encounter and outcome for models spanning $b \in [1.7, 2.6]\,r_H$ for the isothermal (upper panel) and radiative (lower panel) runs. Successful binary formations are coloured in green, failures in orange. The orientation of the first encounter with respect to the motion in the large-scale AGN disc is designated by marker type; circular for prograde flybys and crosses for retrograde. The vertical grey lines correspond with the impact parameters leading to the deepest first periapses in the gasless case near $b = 1.9975, 2.3225\,r_H$. Most successful captures lie within a homogeneous band spanning $b \in [1.85, 2.4]\,r_H$, though there are some failed captures even for relatively deep periapsis near $b \sim 2.1, 2.3\,r_H$. These exceptions are discussed in the main text.

### 4.3.2 Varying impact parameter

We now consider how varying the initial condition $b$, combined with the choice of equation of state, dictates the outcome of binary formation. The $b$ parameter space, along with the effects of varying the ambient disc density, was studied extensively for the isothermal case in HW1. Fig. 10 depicts the dependence of the initial periapsis depth $r_p$ on the BH impact parameter $b$; as discussed in Section 4.3.1 $r_p$ tends to be the defining characteristic that predicts the success or failure of a potential binary formation event. We first note the difference in shape of $r_p = f(b)$ when the equation of state is varied: changes in minidisc structure have perturbed the BHs on to slightly different trajectories pre-encounter, resulting in different periapsis depths during first close encounter. While the double-valley structure observed in the gasless case (Boekholt, Rowan & Kocsis 2023) is clear in the radiative case, in the isothermal case the system has effectively merged into a single wide valley. This is likely due to the strength of gas gravitation on the BHs before close encounter, the isothermal system being more dense (see Section 4.2.1) features a stronger pull on the BHs. The isothermal valley structure is roughly similar to the highest density runs reported in HW1 (fig. 13 therein), though here the ambient density is four times greater and the BH initial conditions are modified (see equation 28).

Generally, successful binary formations lie within the valley structure; models with $b$ values too small perform horseshoe orbits and models with large $b$ values exhibit wide flybys and in either case the BHs do not penetrate deep enough through each other's minidiscs to drive significant dissipation. We note some exceptions to this rule:

(i) a single run with $b = 1.75\,r_H$ in the isothermal system which successfully forms a binary during a relatively shallow encounter

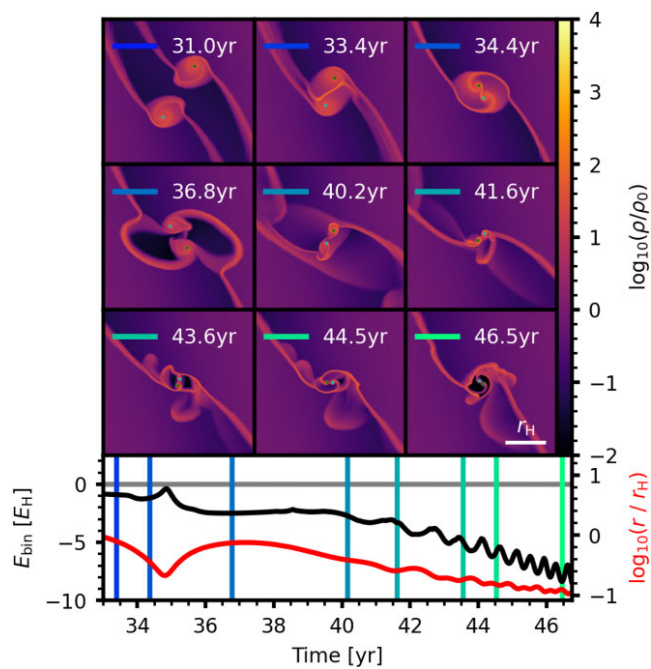

**Figure 11.** Binary formation without a severe close encounter for an isothermal simulation with $b = 1.75\,r_H$. The cyan and green circles represent the smoothing radius $h = 0.05\,r_{H,s}$ about each BH. After dissipating a minor amount of energy in the first encounter, the binary undergoes a steady inspiral with relatively weak dissipation. The binary continues to harden without any extreme close encounters, reaching a stable $E_{bin} = -8\,E_H$ without ever penetrating $r < 0.1\,r_H$. Hardening in this phase is likely to be driven more by large-scale gas geometries such as the trailing circumbinary streamers, as opposed to the impulsive BH-minidisc lag observed in strongly colliding systems. The resulting binary is much more circular ($e \sim 0.1$). See the solid black line in Fig. C1 for the BH trajectories.

(ii) three runs with $b = 2.05, 2.1, 2.35\,r_H$ in the radiative system where binaries fail to form despite experiencing relatively deep encounters

The chronology of the former case is depicted in Fig. 11. This event is similar to fig. 14 in HW1, where an extended dissipation event occurred during a relatively shallow first encounter. In contrast, isothermal runs with slightly deeper first encounters at $b = 1.8, 1.85\,r_H$ did not form a binary. In these two cases, despite a similar amount of energy being dissipated in the first flyby, there was no second encounter; in the $b = 1.75\,r_H$ case it appears to be the SMBH which encourages a second encounter, during which gas dissipation can successfully stabilize the binary. The ability for a third body to drive multiple encounters between quasi-bound binaries is explored thoroughly in Boekholt et al. (2023). As such we have a rare system where the first close encounter is insufficient to form a stable binary, but the post-encounter trajectory is such that a subsequent encounter is able to harden the system sufficiently. It is possible that there are other impact parameters where similar encounters could occur, but they are not sampled in this work's impact parameter space. The ability for binary formation to proceed without severe close encounter implies the existence of radiative binary systems that do not generate blast waves but still form binaries successfully: such a system was not sampled in this work's $b$-space.

In the three radiative systems that fail to form a binary, the BHs still experience strong gas gravitation during close encounter. However, the BH trajectories and gas geometries are such that gas gravitation injects similar amounts of energy on infall as it removes immediately after the first periapsis, resulting in inefficient net dissipation. It is not immediately clear what sets these trajectories apart from those runs which do feature strong net dissipation. Fig. 12 depicts the chronology of such an encounter. As the encounters feature relatively deep periapses, they result in novae generation, but in this case without associated binary formation.

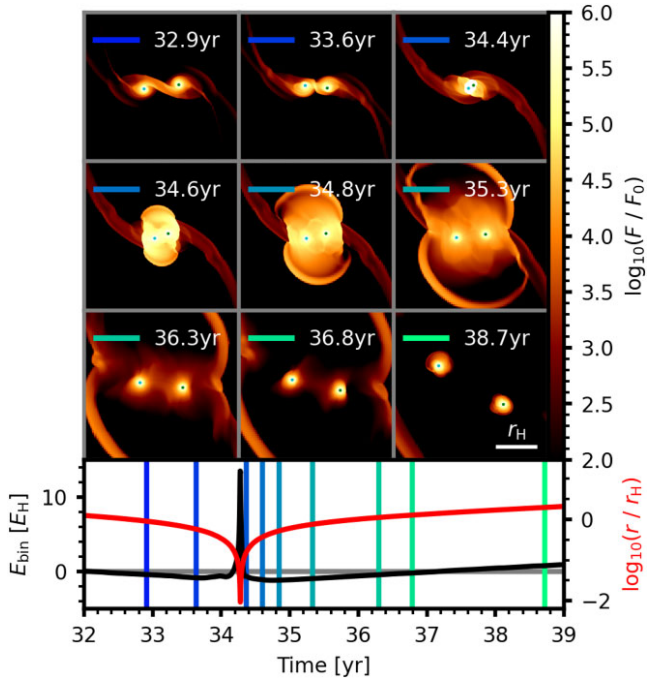

**Figure 12.** Failed binary formation despite severe close encounter for a radiative simulation with $b = 2.1 r_H$. The cyan and green circles represent the smoothing radii $h = 0.05 r_{H,s}$ about each BH. During the close encounter gas gravitation injects as much energy as it dissipates, resulting in almost zero net change. A single strong blast wave is generated during the close encounter, with the BHs flying apart afterwards: no stable binary is formed. The chronology is very similar to the radiative models with $b = 2.05, 2.35 r_H$ that also fail to form a binary despite severe close encounter. See the dashed orange line in Fig. C1 for the BH trajectories.

## 5 DISCUSSION

### 5.1 Comparison to literature

This study builds directly on the work of HW1 that features an extensive comparison of the isothermal system to recent work in the field of embedded binary black hole formation. The effects of radiation on the formation scenario had yet to be considered prior to this study, and so lacks a directly comparable counterpart.

Grishin et al. (2021) considered a system in which a supernova was generated within an AGN disc. While the blast in that study was driven by a different progenitor with higher detonation energies, they offer an insight into the behaviour of a strong shock propagating vertically through the AGN disc (Figs 9 and 10), which we are unable to observe in our 2D study. That study predicted that much of the detonation energy is lost vertically out of the disc (as the path of least resistance), which may lead to decreased disruption of the circumbinary environment and brighter flares as energy is injected into regions of lower optical depth. Future studies into minidisc collisions in 3D may wish to compare to this supernovae system to see what morphological/energetic differences arise.

### 5.2 Caveats and limitations

In our simulations, we adopt several assumptions that should be considered when interpreting the results and may warrant further study.

#### 5.2.1 3D effects

The arrival of strong blast waves in the binary formation system marks the introduction of likely non-planar effect in a 2D simulation. While the pre-encounter system appears to be reasonable well described by evolving the vertically integrated Navier–Stoke's equations, we do not expect the strong shocks generated at periapsis to remain confined to the disc plane. This system clearly requires a 3D treatment if the propagation of these shocks is to be faithfully captured. Consideration of vertical scale heights may also allow for better predictions to the effective local scale height, which in turn can be used to predict the local viscosity. We expect that in 3D, where hot gas material will be able to escape vertically out of the disc to regions of lower optical depth, the flares may be considerably brighter.

#### 5.2.2 Subgrid minidisc physics

While the simulations presented are of sufficiently high resolution to resolve much of the small-scale gas dynamics around the BHs, we are unable to simulate right the way down to the inner edge of the minidiscs, at the innermost stable circular orbit (ISCO). We do not expect this to have a significant effect on the qualitative conclusions concerning the BH dynamics or shock generation, but in failing to simulate down to such small radii we are potentially missing a large section of the minidisc luminosity. If the minidiscs can be modelled as $\alpha$-discs down to arbitrarily small radii, we would expect to reach greater temperatures near the ISCO as $T_{\text{eff}} \sim r^{-\frac{3}{4}}$. This small area, high-temperature region may dominate the total thermal emission of the minidisc, resulting in a hotter, bluer spectrum. We think it likely therefore, that the steady luminosities calculated for the pre-encounter system are underestimates of the true emission.

While it is not immediately clear how the generation of blast waves should change as more of the small-scale minidisc is included in the simulation, we do find that blast waves are persistent in the simulations when an extra level of maximal mesh refinement is included. It seems likely that the amount of gas the minidiscs are able to retain during such encounters, and the exact detonation energies of the blast waves generated, would depend not only on the resolution of the simulation but also the smoothing length of the BH gravitational potential. Regardless of the exact pressure/temperature profiles near the BHs, the generation of large overpressures (and subsequently blast wave propagation) during close encounters should be ubiquitous across severely colliding systems, so we expect the general phenomena documented in this study to be consistent with higher resolution systems.

#### 5.2.3 Accretion

BH accretion is not modelled in this study. This decision arose from an expectation that without a full model of the minidisc flow down to the ISCO, approximations made as to the accretion rate should consider the viscous time-scale $\tau = \frac{r^2}{\nu}$. As discussed in Section 2.1.4, determining $\nu$ self-consistently within the simulation is non-obvious and while reasonable prescriptions have been used in prior studies (Li & Lai 2024), the underlying assumption that viscosity can be predicted using the $\alpha$-disc formula $\nu = \alpha c_s H$ breaks down when physically realistic estimators for $H$ are unavailable. This is the case when the post-periapsis blast waves destroy much of the minidisc structure, as the usual assumptions of vertical hydrostatic equilibrium are clearly invalid. Without a clear means to physically motivate a variable viscosity and therefore accretion rate, we opt to neglect it from this study.

*5.2.4 Gas physics*

More advanced gas effects yet to be considered in the field of embedded binary formation are also neglected from this study:

(i) *BH feedback*: We might expect the BH minidiscs to feed mass, energy, and momentum back into the surrounding gas in the form of jets or winds. These jets may have their own significant non-thermal emission.

(ii) *Gas self-gravity*: While our initial conditions allow us to study binary capture in an environment where the gas mass local to the binary is less than the binary mass, gas self-gravity may still influence the morphology and evolution of the minidiscs. This effect is likely to be more significant for lower mass BHs, or for denser ambient discs.

(iii) *Non-LTE physics*: Our assumption of local thermal equilibrium between the gas and radiation fields is reasonable considering the high optical depths present in the system. It is possible that this may break down within the ultra-hot underdense vacuums driven by periapsis novae, though this was not observed in this study.

(iv) *Magnetism*: This study is purely hydrodynamic, with no consideration of magnetic effects. The introduction of a background magnetic field, with potential strengthening during minidisc formation, may collimate the outburst flow on to field lines. Such a system is likely to favour outflow vertically out of the disc. It is not clear how the introduction of magnetic pressure support might affect the minidisc structure or collision environment.

*5.2.5 Initial conditions*

Of potential interest for future studies would be a consideration of the effects of varying some of the initial conditions, as disc collisions in different environments may result in different detonation energies, luminous chronologies, and spectral emission.

(i) *BH trajectories*: We have launched BHs into the AGN flow co-planar with the disc and with zero eccentricity. Variations in initial inclinations and eccentricity are likely to impact the frequency and depth of close encounters. Dittmann, Dempsey & Li (2024) consider the effect of inclination in pre-existing binaries, but not during formation.

(ii) *Encounter location*: We have studied only a single position within the AGN disc; if estimates are to be made as to formation frequencies across an entire AGN disc, we must consider how capture likelihood changes as a function of position within the disc.

(iii) *AGN parameters*: We have fixed many global parameters that would vary between AGNs, such as the total luminosity, viscosity coefficient, and accretion efficiency. Variations in these quantities would generate different shearing box environments and likely influence capture likelihood and outburst prominence.

Consideration of all possible encounter geometries and environments is key if predictions are to be made of the frequency of embedded BBH formation within cosmic volumes.

# 6 SUMMARY AND CONCLUSIONS

In this work, we have simulated potential binary formation events between initially isolated black holes and compared the effect of changing the equation of state from isothermal to a more realistic adiabatic mixture of gas and radiation. We analysed the morphology of the BH minidiscs prior to close encounter and the resulting likelihood of binary formation. We uncover novel thermal effects introduced by the radiative equation of state, and consider the potential observable consequences of these phenomena. We summarize the key findings as follows:

(i) The radiative minidisc morphology is very different from the isothermal case: larger, more diffuse, and substantially hotter. These minidiscs feature radiation support in their inner regions and are much brighter than the local ambient background. Future studies into BBH formation should definitely consider using a radiative equation of state to faithfully capture these differences.

(ii) Despite their morphological differences, the radiative and isothermal systems dissipate similar amounts of orbital energy by gas gravitation during close encounters (with isothermal systems being slightly more efficient). Both systems conform to the general conclusions of HW1, where deeper close encounters dissipate more energy, and successfully form binaries for a similar range of impact parameters.

(iii) During close encounters in radiative systems, severe collisions between minidiscs generate large overpressures that drive hot, bright fireballs over many Hill radii. These 'disc novae' are 10 times more luminous than their parent minidiscs (and 100 times more than the local ambient gas) but are likely too dim to be visible in the bolometric AGN emission for this collision environment and numerical treatment. Disc collisions in the outer AGN disc may contribute to IR band variability, with $\mathcal{L}_{\text{peak,IR}} \sim 0.03 \mathcal{L}_{\text{AGN,IR}}$. Subsequent periapsis passages in newly formed binaries result in further novae generation. These blasts strip mass from the binary system, damping the effects of gas hardening in the young binary.

(iv) The generation of novae during close encounters represents a new form of BH feedback that may have important effects on the global AGN disc, acting as a significant source of heat. The exact geometry and intensity of this feedback remains ambiguous and requires careful consideration in 3D for further study. It is possible that collisions in different AGN environments and with more realistic physics may be brighter and more prominent against the AGN background.

In simulating BH close encounters within an AGN disc using a radiative equation of state, we show the isothermal prescription to be an inadequate description of local gas hydrodynamics and uncover novel BH feedback phenomena in the form of disc novae.

**ACKNOWLEDGEMENTS**

All simulations presented in this paper were performed on the Hydra cluster at the University of Oxford. This work was supported by the Science and Technology Facilities Council grant number ST/W000903/1. TB's research was supported by an appointment to the NASA Postdoctoral Program at the NASA Ames Research Center, administered by Oak Ridge Associated Universities under contract with NASA.

## DATA AVAILABILITY

The data underlying this article will be shared on reasonable request to the corresponding author.

## APPENDIX A: SMOOTHING RADII

When using a non-isothermal equation of state, adopting the smoothing length $h = 0.01 r_{H,s}$ as implemented in HW1 lead to rapid non-physical heating when the BHs were impulsive introduced to the system at $t = 0$. We believe this to be due to exerting too steep a force gradient across cells close to the BHs resulting in errors in energy flux calculations. It is possible to achieve greater stability with smaller smoothing radii if the resolution near each BH is increased, but doing so significantly increases the computational cost. The adopted value of $h = 0.05 r_{H,s}$ in this work represents the best available compromise between realistic physics and numerical stability.

## APPENDIX B: ROUND-OFF ERRORS

During very close periapsis passages, the inability to track the BH–BH system to arbitrary precision results in round-off errors. These errors manifest as random noisy perturbations from the true value, visible as 'spikes' in the binary energy. Eliminating these errors requires a significant increase in computational precision: the spikes are effectively eliminated when tracking the system with 30 decimal places. After the BHs have passed through a deep periapsis passage, the perturbations rapidly decrease. While the magnitude of the spikes can be large, the randomness of the noise means that the net effect is insignificant. As such, we opt to continue to use double precision (15 decimal places) throughout this study, as the errors have no effect on our analysis or conclusions.

## APPENDIX C: BLACK HOLE TRAJECTORIES

To better visualize the BH trajectories during encounters, we include Fig. C1 which displays the trajectories of three runs highlighted in the paper. Each line traces the trajectory of the outer BH in the frame of the inner BH. In solid orange, the fiducial radiative run with $b = 2.2r_H$, as presented in Figs 3 and 4. In this run, the BHs dissipated sufficient energy during the first deep encounter to form a stable eccentric binary which rapidly hardens. In solid black, the isothermal run with $b = 1.75r_H$, as presented in Fig. 11. In this run, despite the relatively shallow first encounter, the BHs are still able to form a stable (but relatively soft), low-eccentricity binary. In dashed orange, the radiative run with $b = 2.1r_H$, as presented in Fig. 12. While this run features a deep first close encounter, insufficient dissipation results in no binary formation.

This paper has been typeset from a T<sub>E</sub>X/LAT<sub>E</sub>X file prepared by the author.

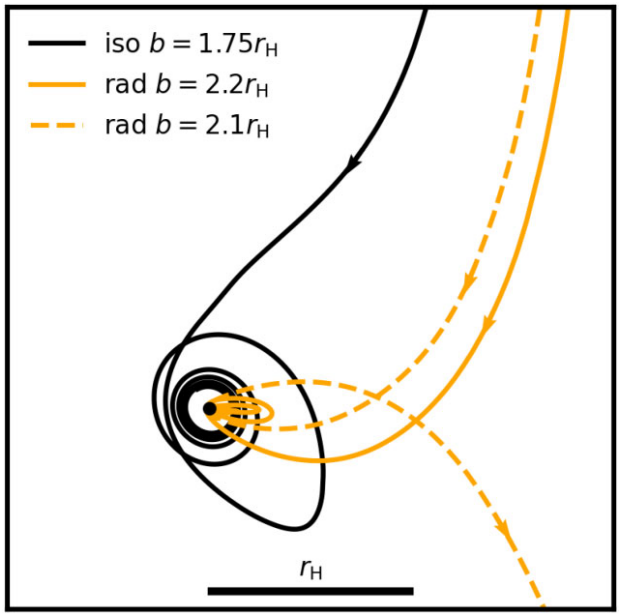

**Figure C1.** The path of the outer BH in the frame of the inner BH, for three of the runs discussed in the paper. The outcome of each run is very different, with the initially unbound BHs forming either a high-eccentricity retrograde binary (solid orange), low-eccentricity prograde binary (solid black), or no binary at all (dashed orange).